\documentstyle[prb,aps,twocolumn]{revtex}

\draft

\begin{document}
\twocolumn[\hsize\textwidth\columnwidth\hsize\csname @twocolumnfalse\endcsname
\title{\hfill {\small ITP Preprint Number NSF-ITP-96-61}  \\
\vspace{10pt}
The $\nu=1/2$ Quantum Hall Effect and Symmetry Properties of 
Two-Component Paired Quantum Hall States}
\author{A. M. Tikofsky}
\address{Institute for Theoretical Physics, University of California,
Santa Barbara, CA 93106}
\date{22 July 1996}
\maketitle
\begin{abstract}
We describe paired quantum Hall states at filling fractions $\nu$, where
$\nu^{-1}$ is an integer, and present explicit wavefunctions for these
states.
Experiments are proposed to
distinguish between paired states of singlet and triplet symmetry.
It is argued from existing experimental data that the $\nu=1/2$ state
observed in double layer quantum wells is a singlet and not
a triplet paired state.  In contrast, this state
has been interpreted as a `3-3-1 state' 
which is a triplet paired state.
\end{abstract}
]
\narrowtext

There has been a great deal of interest in the two-component
quantum Hall effect (QHE) as a description of both spin-unpolarized
systems as well as the spatially
separate charge distributions that arise in quantum 
wells.\cite{Books,review}
The potentially rich internal structure of two-component QHE states
has manifested itself in the experimental observation,
in both wide quantum wells
and double quantum wells,\cite{onehalf,gapdata,Pepper}
of the QHE at filling fractions for which the effect is absent
in single-component systems.
This internal structure may be thought of as a pseudospin degree of freedom
such that electrons in the top and bottom
layers have up and down pseudospin, respectively.
Using this convention, we classify paired quantum Hall 
states according to whether the pseudospin component of their
two-particle pair wavefunction has triplet or singlet symmetry.
Experimental probes that distinguish between triplet and
singlet pairing will be discussed and the corresponding
arguments will be applied to
the $\nu=1/2$ QHE.  The accepted explanation of this state
is given by the `3-3-1' paired quantum Hall state, one of a manifold
of pseudospin-triplet paired states at $\nu=1/2$.\cite{331,SongHe,Ho-331}
However, we will argue that existing experimental studies of the
$\nu=1/2$ QHE are inconsistent with a triplet paired state and suggest 
reinterpreting the $\nu=1/2$ QHE as a singlet paired
state.\cite{gapdata,Manoharan}

We distinguish between
a triplet and singlet pseudospin two-particle pair wavefunction,
a quantity which will be precisely defined later in the paper.
In the ground state, the pseudospin component $T$ of the two-particle 
pair wavefunction is assumed to
be the same for all pairs, constant throughout space, and
either a singlet (s) or triplet (t) state where
\begin{eqnarray}
\label{pseudospin}
T^s_{j,k}&=&{1\over\sqrt{2}}(\uparrow_j\downarrow_k-
\downarrow_j\uparrow_k)\nonumber\\
T^t_{j,k}(\vec{a})&=&{a_o\over \sqrt{2}}
(\uparrow_j\downarrow_k+\downarrow_j\uparrow_k)
+a_u \uparrow_j\uparrow_k+a_d\downarrow_j\downarrow_k\ ,
\end{eqnarray}
and $\vec{a}=(a_o,a_u,a_d)$ is any unit length complex 3-vector.
The manifold of triplet states at a given filling
fraction enable us to adiabatically change the triplet pair
pseudospin wavefunction $T^t$ and preserve the structure of the
triplet paired state.  The singlet state however is unique and
its pseudospin degrees of freedom are rigid.
In this work, we will contrast the rigidity of singlet paired states
with the freedom of triplet paired states to adjust pseudospin degrees
of freedom in order to minimize pseudospin energy.
We will further
argue that the pseudospin degrees of freedom can have a 
dramatic influence on the energy gap of the QHE state.

Let us construct a phenomenological Hamiltonian for the pseudospin
degrees of freedom in order to understand their dynamics as well
as their influence on the QHE energy gap.
Since we assume an ordered state with an associated pseudospin 
order parameter $T$, we can truncate our phenomenological
Landau-Ginzburg Hamiltonian density at quadratic 
order in the local density $S_i({\bf r})$
of the $i^{\rm th}$ component of pseudospin.  Because we
are interested in the ground state, we do not include gradient
terms which reflect the energy cost of spatially varying the spin
density.
The pseudospin Hamiltonian is assumed to preserve the relative
densities of the two layers and to be invariant
under overall spin-flip $\uparrow\iff\downarrow$.
To quadratic order, the most general Hamiltonian 
density satisfying these symmetries is
$h_{int}({\bf r})=c_\parallel S_\parallel^2({\bf r})
+c_z S_z^2({\bf r})$ where $S_\parallel^2=S_x^2+S_y^2$.
The physical interaction Hamiltonian includes only repulsive density-density
interactions and therefore $c_z>0$ and $c_\parallel=0$.
For instance, the momentum space pseudospin dependent interaction
for two layers spaced a distance $d$ apart is
$V_{\sigma,\sigma'}={2\pi e^2 \over q}[\delta_{\sigma,\sigma'}+
e^{-qd}(1-\delta_{\sigma,\sigma'})]$ and the corresponding
value of $c_z=4 \pi e^2 (1-e^{-q d})/q\simeq 4\pi e^2 d$ while 
$c_\parallel=0$.  We therefore assume that
$0\le c_\parallel<< c_z$ which implies
that the lowest energy pseudospin pair wavefunction lies in the
$x-y$ plane.  In addition to the interaction term, we account for
a relative gate potential $U$ between the two layers and
a matrix element $t$ for tunneling between these layers in
the total pseudospin Hamiltonian density 
\begin{equation}
\label{hpseudo}
h({\bf r})=c_\parallel S_\parallel^2({\bf r})
+c_z S_z^2({\bf r}) -2tS_x({\bf r}) -U S_z({\bf r})\ .
\end{equation}
In studies of two-component quantum Hall systems, it
is usually assumed that $c_\parallel=0$ and therefore,
when $t=U=0$, the Hamiltonian breaks the SU(2) pseudospin
symmetry down to U(1) but still has a gapless ferromagnetic
`spinwave mode' in the {\it triplet} pseudospin degrees of
freedom.  A nonzero $t$ or $U$ breaks the U(1) symmetry
and gives this mode a gap $\Delta_{spin}$.

{\bf From the effect of an applied gate voltage $U$ on the energy
gap of a paired QHE state, we can distinguish between a
triplet or singlet paired state.}
The triplet state can accommodate an applied voltage by
modifying the pseudospin wavefunction
$T^t$ in order to change the relative density of electrons in the two
layers.  For example, taking $|a_u|>>|a_o|>>|a_d|$ 
in Eq. (\ref{pseudospin}) implies significant charge imbalance between
the two layers.  In analogy with the behavior of quantum Hall ferromagnets
and their associated Skyrmionic excitations,\cite{review,Skyrmions}
a triplet state takes advantage of the applied voltage to increase
its energy gap because the voltage term $-US^z$ in Eq. (\ref{hpseudo})
acts like a Zeeman term in pseudospin space.
In contrast, a singlet state by definition has equal number of electrons
in both layers, can not change the distribution of charge
to accommodate the external voltage $U$, and will lose
energy because of the applied external voltage $U$. 
The energy gap for a singlet state should vanish when the amount
of energy gained by redistributing charge exceeds the zero voltage
energy gap of the state.
When we compare the
zero voltage gap of a singlet paired quantum Hall state to the 
gain in capacitive energy from
redistributing charge among the two layers in response to
an applied voltage $V\sim U$, we expect the QHE gap to vanish
experimentally when
\begin{equation}
\label{condition}
\rho\Delta_{1/2}(V=0)\sim CV^2
\end{equation}
where $C$ is the capacitance per unit area of the two layer system 
and $\rho$ is the electron density.
{\bf As expected for a singlet state,
the $\nu=1/2$ QHE gap vanishes when the condition in Eq.(\ref{condition}) is
satisfied.}\cite{gapdata,Manoharan}
The $\nu=3/2$ state is believed to be similar in origin
to the $\nu=1/2$ state and also exhibits rapid vanishing of the gap
with applied voltage.
In contrast, the gap of the $\nu=1$ state,
which is interpreted as a quantum Hall ferromagnet and hence a triplet,
does not vanish rapidly as a 
function of $V$.\cite{gapdata,Pepper,Manoharan}
We note that there are two distinct energy scales:
the transport QHE gap $\Delta_{QHE}$ and the potential gain in
pseudospin energy not available to a singlet state $\Delta_{spin}$.
The observed QHE gap for a singlet state
$\Delta_{QHE}(U,t)\sim\Delta_{QHE}(0,0)-\Delta_{spin}$ 
decreases with increasing $\Delta_{spin}$ and vanishes for
sufficiently large $\Delta_{spin}$.
When the unreasonable approximation is made of ignoring the
terms quadratic in $S_i^2$ in Eq.(\ref{hpseudo}),
we get the heuristic estimate $\Delta_{spin}\simeq\sqrt{4t^2+U^2}$.

The preference of the physical system for singlet or triplet pairing is
determined by the full Hamiltonian and not merely the
pseudospin Hamiltonian in Eq.(\ref{hpseudo}).
We have studied the pseudospin degrees of freedom on the
premise that the full interaction Hamiltonian has a unique 
paired QHE ground state with a gap to excitations.
Despite the experimental evidence, we
need to address numerical studies of the full interaction Hamiltonian
in a wide quantum well.
These studies, using a basis set of states projected onto the
lowest Landau level, found a unique ground state with a gap
to excitations at $\nu=1/2$ 
for intermediate values of the layer spacing 
$d$ but not when $d=0$ or $d=\infty$.\cite{SongHe}  
In addition, these studies of very small systems found
a large overlap with the triplet paired `3-3-1' state
with $T^t=T^{331}=\uparrow\downarrow+\downarrow\uparrow.$
Because this is a finite size study, there can also be a
a large overlap with a candidate singlet paired state.
This is possible because the spatial dependence of the two-particle
pair wavefunction implies that the overlap between
a triplet paired state and a singlet paired state need only vanish
in the thermodynamic limit.  
Another argument stated in favor of the `3-3-1 state' is that,
in both experimental data and finite-size numerical calculations,
the QHE gap is stable only for intermediate values of the layer spacing $d$.
Similar behavior is plausible for a singlet paired state.
Further numerical studies are necessary in order
to compare singlet and triplet paired states energetically.

Let us now describe fully antisymmetrized wavefunctions for two-component
paired quantum Hall states
including the `3-3-1 state.'\cite{Ho-331,Herbut,Wilczek,Read,Pfaffian}
The paired wavefunction
\begin{equation}
\label{wfn}
\Psi_{2n}=\prod_{i<j}(z_i-z_j)^{2n} Pf\{ G_{k,l}\} 
e^{-{1\over 4}\sum_{i}|z_i|^2}
\end{equation}
is a properly antisymmetrized function of the electron coordinates
$z_i$ at filling fraction $\nu=1/{2n}$ where {\it the two-particle pair
wavefunction}
\begin{equation}
\label{wfng}
G_{k,l}=g(z_k-z_l)\ T_{k,l}
\end{equation}
is an antisymmetric function of its coordinates and
$g$ and $T$ are functions of only
the spatial coordinates and pseudospin coordinates, respectively.
We construct singlet or triplet paired states by
taking $G^s=g^s T^s$ or $G^t=g^t T^t$, respectively, where
the $T$'s are defined in Eq. (\ref{pseudospin}), and
\begin{equation}
\label{wfna}
g^s(z_k,z_l)={f(|z_k-z_l|)\over (z_k-z_l)^{2p}}\qquad
g^t(z_k,z_l)={f(|z_k-z_l|)\over (z_k-z_l)^{2p+1}}\ .
\end{equation}
T. L. Ho has recently shown,\cite{Ho-331}
that upon antisymmetrization, the `3-3-1 state' is
a triplet paired state given by $\Psi_{2n}$ with $n=1$,
\begin{equation}
\label{wfn331}
T^{331}_{k,l}=\uparrow_k\downarrow_l+\downarrow_k\uparrow_l \ ,\ {\rm and}
\qquad g^{331}_{k,l}={1\over {z_k-z_l}}\ .
\end{equation}
As was claimed, the `3-3-1 state' is one of a manifold of
states with different pseudospin pair wavefunctions
$T^t$ that can be adiabatically
deformed into each other via a rotation in pseudospin space.
We can also define paired states at any filling fraction for
which a singlet component QHE state exists.
Given a conventional single component quantum Hall state $\phi$ at
filling $\nu$, 
\begin{equation}
\label{wfn2}
\Phi=\phi Per\{ H_{k,l}\}
\end{equation}
is a properly antisymmetrized pair wavefunction
\cite{Herbut,Wilczek,Read,Pfaffian} at filling $\nu$ where 
\begin{equation}
\label{wfnHH}
H_{k,l}=h(z_k-z_l)\ T_{k,l}
\end{equation}
is a symmetric function of its coordinates and
$h$ and $T$ are functions of only
the spatial coordinates and pseudospin coordinates, respectively.
Because $H$ is a symmetric function, we construct singlet or
triplet paired states by 
taking $H^s=h^s T^s$ or $H^t=h^t T^t$, respectively, where
the $T$'s are defined in Eq. (\ref{pseudospin}) and
\begin{equation}
\label{wfnh}
h^t(z_k,z_l)={f(|z_k-z_l|)\over (z_k-z_l)^{2p}}\qquad
h^s(z_k,z_l)={f(|z_k-z_l|)\over (z_k-z_l)^{2p+1}}\ .
\end{equation}
A state with $h^t=1$ is just the
conventional quantum Hall ferromagnet at filling fraction $\nu$.

In defining the nature of a paired state, we distinguish
between a two-particle
pair wavefunction that decays as a power law with distance
and one that decays exponentially.\cite{tikof-kiv}
When the two-particle pair wavefunction is short-ranged, 
the bosonic pairs have a well-defined size and the long-wavelength
behavior is accurately described in terms of these
pairs.  In contrast, a two-particle pair wavefunction 
that decays as a power-law has inherently long-ranged pairing.
Because the bosonic pairs do not have a definite size,
the long-wavelength physics of this state can not
simply be described in terms of these pairs. 
We distinguish between short-range and long-range pairing via
the function $f(|z_k-z_l|)$ in Eq.s (\ref{wfna},\ref{wfnh}).
Choosing $f$ to decay exponentially with distance
implies short-range pairing.  
On the other hand, if $f$ is assumed to be constant then the
pair wavefunction in Eq.(\ref{wfn}) is analytic and resides only in the lowest 
Landau level as long as the two-particle pair
wavefunction $G$ decays as a power law such that 
$g(z)\sim z^{-m}$ and $2n\geq m$.  The `3-3-1 state' described
in Eq.(\ref{wfn331}) has such a wavefunction.
In contrast to an analytic power-law
paired state, a short-range paired state is
nonanalytic in such a way that it has nonzero occupation
of all Landau levels and hence costs a finite amount of kinetic energy.
However, the case has recently been made that the loss of kinetic energy
associated with short-range pairing
may be outweighed by a gain in interaction energy for low
magnetic fields.\cite{tikof-kiv,pikus-tikof}

Let us consider both power-law and short-range singlet paired states
at $\nu=1/2$.
Haldane and Rezayi proposed a $\nu=1/2$ power-law singlet paired state
in the lowest Landau level given by Eq.s (\ref{wfn}-\ref{wfna}) with
$n=1$ and $g^s(z_k,z_l)=1/(z_k-z_l)^2$.\cite{Haldane-Rezayi}
The Haldane-Rezayi 
state includes Jastrow factors that keep electrons not in the same pair
well separated but has no Jastrow factors keeping apart electrons in the same
pair.  This state is therefore believed to be stable for 
so-called `hollow-core' potentials that are small at the origin.
Because a short-ranged paired wavefunction need not be analytic,
we can construct a singlet paired state at $\nu=1/2$ that keeps 
electrons in the same pair well separated by choosing a
pair wavefunction $g^s$ that is short-ranged and yet remains finite at the
origin.  Therefore, a short-range paired state can have
correlations between
up and down spins that are absent in the Haldane-Rezayi state.
If these correlations 
gain enough energy for the state then
a short-range singlet paired state could have a much lower
interaction energy than the Haldane-Rezayi state and be experimentally
stable in small to moderate magnetic fields.

The topology of the experimental QHE phase diagram can provide evidence for
short-range pairing when direct continuous phase transitions,
forbidden for conventional QHE states, are observed.\cite{tikof-kiv}  
We first note that the exponentially paired states
described in Eq. (\ref{wfn}) at filling fraction 
$\nu=1/2n$ have the long-wavelength physics of a bosonic Laughlin
QHE state of charge $2e$ bosons at bosonic filling
fraction $\nu_b=\nu/4$. 
The standard picture of QHE phase transitions tells
us that a bosonic Laughlin state at filling fraction $\nu_b=1/8n$ 
has a single conventional QHE edge mode and can therefore make
a direct continuous transition to an insulator.\cite{Books}
In contrast, both the `3-3-1' and the Haldane-Rezayi state
have two distinct edge modes and can evolve
continuously to an insulator only via two separate continuous 
transitions.\cite{331,tikof-kiv,Read2}  The experimental observation
of a direct continuous transition from $\nu=1/2$ to an
insulator would provide evidence for an exponentially paired state.
The topology of the phase diagram can also be employed to make
statements about pairing states of the type described in 
Eq.s (\ref{wfn2}-\ref{wfnh}).
With $h^t=1$ in Eq. (\ref{wfnh}),
the wavefunction $\Phi$ at $\nu=1/(2n+1)$ describes the usual quantum
Hall ferromagnetic states that can make a direct continuous
transition to an insulator.  Short-range paired
states at these filling can also undergo this transition.
The ability of a two-component state at filling $1/(2n+1)$
to make a direct continuous transition to an insulator would only rule out
singlet power-law paired states, and triplet power-law paired
states with $h^t\sim z^{-p}$ ($p\neq 0$).

Inelastic light scattering may provide an experimental probe
to distinguish between triplet and singlet pairing.
Light with a momentum in the direction perpendicular to
the planes $q_z\sim\pi/d$, where $d$ is the distance between
the planes, couples directly to the pseudospin mode in which electrons
in opposite layers move in opposite directions.\cite{MacD}
A strong absorption peak is expected for a triplet state at
the frequency $\Delta_{spin}/\hbar$
because the light directly couples to the triplet pseudospin excitation.
Furthermore, $\Delta_{spin}$ can be varied by tuning the gate voltage $U$.
While a singlet power-law paired state should not exhibit
resonant absorption exactly at $\nu=1/2$, it will have excitations
with nonzero pseudospin whose density will increase as
$|\rho-eB/2hc|$, the deviation from $\nu=1/2$.
Given a disordered system, the absorption spectrum for 
a singlet power-law paired state should 
have a broad ill-defined peak.
In contrast, a singlet short-range paired state would
only have singlet excitations and should not even display a broad
peak in its absorption spectrum.

The author has benefited from useful conversations with
Song He, D. E. Khmelnitski, S. A. Kivelson, A. H. MacDonald,
H. C. Manoharan, M. S. Sherwin, and S. L. Sondhi.
This work has been supported by the National Science Foundation
under grant PHY94-07194
at the Institute for Theoretical Physics.


\begin{references}
\bibitem{Pfaffian}
Given an order $2N$ matrix M$_{ij}$, the permanent of a symmetric matrix
$M$ is Per$(M)=H_1(M)$ and the Pfaffian of an antisymmetric
matrix $M$ is Pf$(M)=H_{-1}(M)$ where 
$H_\alpha(M)=\sum_{P}(\alpha)^{S\{P\}}\prod_{ab}M_{ab}$,
the product is over the $N$ pairs ${ab}$, the sum is over all 
possible permutations of these pairs, and $S\{P\}$ is $-1$
or $+1$ if the permutation $P$ is odd or even, respectively.
\end{references}
\end{document}